\documentclass[10pt,a4paper]{article}
\usepackage[tbtags]{amsmath}
\usepackage{amssymb}
\usepackage{graphicx}
\usepackage[latin1]{inputenc}

\usepackage{amsbsy,amsfonts,fixmath,isomath,theorem}

\usepackage{pgf,pgfarrows,pgfnodes,tikz,color,tkz-berge}
\usetikzlibrary{arrows,shapes,snakes,automata,backgrounds,petri,topaths,calc}

\definecolor{back}{RGB}{249,242,215}

\numberwithin{equation}{section}

\DeclareMathOperator*{\argmin}{\arg\!\min}

\newcommand{\vect}[1]{\ensuremath{\mathbold{#1} } }

\newcommand{\R}{{\mathbb R}}
\newcommand{\C}{{\mathbb C}}

\newcommand{\D}[2]{ \ensuremath{ \frac{\mathrm{d} #1 }{\mathrm{d} #2 } }}

\newtheorem{theorem}{Theorem}[section]

\theoremstyle{definition}

\theoremstyle{remark}

\begin{document}

\title{Model reduction of biochemical reactions networks
by tropical analysis methods}


\author{Ovidiu Radulescu$^1$, Sergei Vakulenko$^2$, and Dima Grigoriev$^3$,  \\
\small  $^1$ DIMNP UMR CNRS 5235, University of Montpellier 2, Montpellier, France, \\
\small  $^2$ University of Technology and Design, St.Petersburg, Russia, \\
\small  $^3$ CNRS, Math\'ematiques, Universit\'e de Lille, 59655, Villeneuve d'Ascq, France.
 }

\maketitle


\centerline{\bf Abstract}

We discuss a method of approximate model reduction for networks of biochemical reactions.
This method can be applied to networks with polynomial or rational reaction rates
and whose parameters are given by their orders of magnitude. In order to obtain
reduced models we solve the problem of tropical equilibration that is a system of
equations in max-plus algebra.
In the case of networks with fast nonlinear cycles we have to compute the
tropical equilibrations at least twice, once for the initial system and a second
time for an extended system obtained by adding
to the full system the differential equations satisfied by the conservation laws of the
fast subsystem.
Our method can be used for formal model reduction in
computational systems biology.


{\bf AMS subjects:} 92C42,  14T05, 34K19.

\section{Introduction}

Networks of chemical reactions are widely used in chemistry
for modeling catalysis, combustion, chemical reactors,
or in biology as models of signaling, metabolism, and gene
regulation. In order to cope with growing amounts of data, these models tend to be
as comprehensive as possible by integrating many variables and processes with several
different timescales.
For most problems in computation and analysis of complex systems, the
upper limit on the size of the system that can be studied has been reached.
This limit can be very low, namely tens of variables for system identification,
symbolic calculation or bifurcation of attractors of large dynamical systems.
Model reduction is a way to bypass these limitations by replacing
large scale models with ones containing less parameters and variables, that are
easier to analyse.

There are several traditional numerical methods for reducing networks of chemical reactions.
 These methods, such as computational singular perturbation (CSP, \cite{lam1994csp}), intrinsic low dimensional  manifold (ILDM, \cite{maas1992simplifying}) exploit the separation of timescales of various processes and variables of the model.
 In dissipative systems, fast variables relax  rapidly to some low dimensional attractive
 manifold called invariant manifold \cite{gorban2005invariant} that carries the slow mode  dynamics. A projection of dynamical equations
 onto this manifold provides the reduced dynamics \cite{maas1992simplifying}.

In the last decade, the rapidly growing field of
 computational and systems biology produced biochemical reactions networks models for cell physiology, of increasing size and complexity. Model reduction has been identified as a highest-priority challenge for these fields, expected to tame the complexity  and
 simplify the analysis of biological models. However, these models came with peculiarities and
the extant traditional model reduction methods are not entirely suitable for this endeavour.
Biological models suffer from structural and parametric uncertainty and one rarely has
precise information about kinetic parameters.
One of the main problem of
computational biology is the parameter
space exploration and analysis of possible model behaviors.
Therefore, formal, symbolic, or semi-quantitative model reduction methods are more
appropriate than numerical methods that need precise
parameters.

Formal model reduction can be based on conservation laws, exact lumping \cite{Feret21042009},
and more generally, symmetry \cite{clarke1996exploiting}. 
For chemical reactions networks with fast reaction cycles and fast species, lowest
order approximations of attractive invariant manifolds are provided by  quasi-equilibrium
or quasi-steady state approximations \cite{GRZ10ces}.
These two approximations allow model reduction by  graph reconstruction via linear lumping,
pruning, and algebraic elimination of fast variables \cite{radulescu2008robust,radulescu2012frontiers}.
Graphical reduction methods use elementary modes \cite{Clarke92}, or the Laplacian defined on the
graph of complexes of the reactions network \cite{rao2013graph},
 but have little or no connection with singular
perturbation methods and do not exploit multi-scaleness of biochemical networks.
A fully formal reduction method exploiting orders of magnitude of
variables and parameters  is still missing.

In this paper we present a new model reduction method, inspired by tropical geometry and
analysis. This method is particularly suited for computational biology because it
combines graphical approaches, semi-quantitative reasoning and symbolic manipulation.

The plan of the paper is the following. The second section introduces the
biochemical reactions models and the tropical geometry concepts
needed for the presentation of our results. In the third section we discuss the relation between
tropical variety and Newton-Puiseux series. In the fourth section we provide general results
of existence of an invariant manifold for biochemical systems with fast species and fast cycles.
We also discuss how to choose slow and fast variables in this case, and how to define reduced models describing
the slow dynamics on the invariant manifold and the fast relaxation towards this manifold.
In the fifth section we apply these results to a nonlinear cycle of reactions.

\section{Definitions and settings}
We consider biochemical networks described by mass action kinetics
 \begin{equation}
 \D{x_i}{t} = \sum_j k_j S_{ij}  \vect{x}^{\vect{\alpha_{j}}}.
 \label{massaction}
 \end{equation}
where $k_j >0$ are kinetic constants, $S_{ij}$ are the
entries of the stoichiometric matrix (uniformly bounded integers,
$|S_{ij}| < s$, $s$ is small),
$\vect{\alpha}_{j} = (\alpha_1^j,\alpha_2^j, \ldots, \alpha_n^j)$ are multi-indices,
  and $\vect{x^{\vect{\alpha}_{j}}}  = x_1^{\alpha_1^j} x_2^{\alpha_2^j} \ldots x_n^{\alpha_n^j}$.
 We consider that  $\alpha_i^j$ are positive integers. However, the approach can be extended to rational or
real indices.

The choice \eqref{massaction} is not restrictive, because most kinetic
laws used in computational biology can be decomposed into simpler steps each one obeying
mass action law \cite{King,Temkin65}. Extensions of our approach, directly applicable to
models whose rate functions are ratios of two polynomials (such as Michaelis-Menten or Hill functions)
without expanding them into mass action elementary steps, were briefly
discussed in \cite{NGVR12sasb} and will be presented in detail in future work.
S-systems, used to model metabolic networks and for which $\alpha_{j}$ are rational
or real multi-indices \cite{savageau1987recasting}, are also
covered by our approach.

For slow/fast systems, the slow invariant manifold
is approximated by a system of polynomial equations for the fast species.
 This crucial point allows us to find a connection with tropical geometry. We introduce now the
terminology of tropical geometry needed for the presentation of our results, and refer to
\cite{maclagan2009introduction} for a comprehensive introduction to this field.

Let $f_1, f_2, \ldots, f_k$ be  multivariate polynomials, $f_i \in \C[x_1, x_2, \ldots,x_n]$.
By considering sums of products of these polynomials
by arbitrary polynomials
we define the ideal
$I \subset \C[x_1, x_2, \ldots,x_n]$
generated by them. The ideal is important in the context of solving systems of algebraic equations
because any solution of the system
$f_1(\vect{x})=0,f_2(\vect{x})=0, \ldots, f_k(\vect{x})=0$
is also a solution of $f(\vect{x})=0$ where $f \in I$. Important reasons
for considering the generated ideal in the context of model reduction
will be discussed in the Section \ref{newton}.

Let us now consider that variables $x_i,\, i \in [1,n]$ are written as powers of a small positive
parameter $\epsilon$, namely $x_i = \bar{x}_i \varepsilon^{a_i}$, where $\bar{x}_i$ has order
zero. The orders $a_i$ indicate the
order of magnitude of $x_i$. Because $\epsilon$ was chosen small, $a_i$ are lower for larger
absolute values of $x_i$. Furthermore, the order of magnitude of monomials $\vect{x^{\vect{\alpha}}}$
is given by the dot product  $\mu = <\vect{\alpha},\vect{a}>$, where $\vect{a} = (a_1,a_2,\ldots,a_n)$.
Again, smaller values of $\mu$ correspond to monomials with larger absolute values.
For each multivariate polynomial $f$ we define the tropical hypersurface $T(f)$ as the set of vectors
$\vect{a} \in \R^n$ such that the minimum of $<\vect{\alpha},\vect{a}>$ over all monomials in $f$
is attained for at least two
monomials in $f$. In other words, $f$ has at least two dominating monomials.

A {\em tropical prevariety} is defined as the intersection of a finite number of tropical
hypersurfaces, namely $T(f_1,f_2,\ldots,f_k) = \cap_{i\in[1,k]} T(f_i)$.

A {\em tropical variety} is the intersection of all tropical hypersurfaces in the ideal
$I$ generated by the polynomials $f_1,f_2,\ldots,f_k$, namely
$T(I) =  \cap_{f \in I} T(f)$.
The tropical variety is within the tropical prevariety,
but the reciprocal property is not always true.

For our purposes, we slightly modify the classical notion of tropical prevariety.
A {\em tropical equilibration} is defined as a vector  $\vect{a} \in \R^n$ such that
$<\vect{\alpha},\vect{a}>$ attains its minimum value
at least twice for monomials of different signs, for
each polynomial in the system $f_1,f_2,\ldots,f_k$. Thus, tropical equilibrations are subsets of
the tropical prevariety. Our sign condition is needed because we are interested in approximating
real positive solutions of polynomial systems (the sum of several dominant
monomials of the same sign have no real strictly positive roots).

In this paper we discuss how tropical equilibrations can be used for model reduction of chemical reactions networks. Tropical equilibrations indicate dominant monomials whose equality define approximated invariant manifolds. Furthermore, they can be used to compute the timescales of the
species, which is important for deciding which species are fast and
can be eliminated by quasi-stationarity conditions.

More precisely, we assume that parameters of the kinetic models
\eqref{massaction} can be written as
\begin{equation}
k_j = \bar k_j \varepsilon^{\gamma_j}. 
\label{scaleparam}
\end{equation}
The exponents $\gamma_j$ are considered to be integer or rational. For instance, the following
approximation produces integer exponents:
\begin{equation}
\gamma_j = \text{round}( \log(k_j) / \log(\varepsilon)),
\end{equation}
where  round stands for the closest integer (with half-integers rounded to even numbers).
Without rounding to the closest integer, changing the parameter $\epsilon$
should not introduce variations in the output of our method. Indeed, the
tropical prevariety is independent on the choice of $\epsilon$. However,
rounding to integer or rational exponents is needed in order to
ensure that our lowest order approximations can be extended to
series.


Timescales of nonlinear systems depend not only on parameters but also
on species concentrations, which are a priori unknown.
We introduce the species orders vector
$\vect{a} = (a_1,a_2,\ldots,a_n)$, such that $\vect{x} = \bar{\vect{x}} \varepsilon^{\vect{a}}$.
 Of course, species orders vary
 in the concentration space and have to be calculated.
To this aim, the network dynamics is first described by a rescaled ODE system
 \begin{equation}
 \D{\bar{x}_i}{t} = \sum_j \varepsilon^{\mu_j - a_i} \bar k_j S_{ij}   {\bar{\vect{x}}}^{\vect{\alpha_{j}}},
 \label{massactionrescaled}
 \end{equation}
where
\begin{equation}
\mu_j = \gamma_j +  \langle \vect{a},\vect{\alpha_j}\rangle,
\label{muj}
\end{equation}
and $\langle , \rangle $ stands for the dot product. 

The r.h.s.\ of each equation in
\eqref{massactionrescaled} is a sum of multivariate monomials in the concentrations.
The orders $\mu_j$ indicate how large are these monomials, in absolute value.
A monomial of order $\mu_j$ dominates another monomial of order
$\mu_{j'}$ if
 $\mu_j < \mu_{j'}$.

The timescale of a variable $x_i$ is given by $ \frac{1}{x_i}\D{x_i}{t} = \frac{1}{\bar{x}_i}\D{\bar{x}_i}{t}$
whose order is:
\begin{equation}
\nu_{i}  = \min \{ \mu_j |  S_{ij} \neq 0 \} - a_i
\end{equation}
The order  $\nu_{i}$ indicates how fast is the variable $x_i$ (if
$\nu_{i'} < \nu_{i}$ then $x_{i'}$ is faster than $\nu_{i}$) .

{\em The tropical equilibration problem} consists in finding a species order vector $\vect{a}$ such that
\begin{equation}
\min_{j,S_{ij} >0} ( \gamma_j + \langle \vect{a},\vect{\alpha_j}\rangle ) =
\min_{j,S_{ij} <0} ( \gamma_j + \langle \vect{a},\vect{\alpha_j}\rangle )
\label{eq:minplus}
\end{equation}
We have shown in \cite{Noel2013a}
that species orders $a_i$ can be computed as solutions of \eqref{eq:minplus}. As discussed above,
these solutions belong to the tropical prevariety of the polynomials defining the
chemical kinetics. One of the problem of this approach is that the tropical prevariety is too
large, namely one can find too many tropical equilibrations. Although all these equilibrations
can formally lead to reduced models, some are spurious. In this paper we propose to
use the tropical equilibrations in
a smaller set of the tropical variety.
This choice is natural,
because by a result of Speyer and Sturmfels
\cite{speyer2004tropical}
the tropical variety is related to Newton-Puiseux series. As a matter of fact,
choosing tropical solutions in the tropical variety ensures their lifting
to series. Furthermore, using the tropical variety
allows us to overcome
another limitation of our previous application of tropical geometry ideas to model
reduction. Namely, in  \cite{Noel2013a} our reduced models were obtained by tropical truncation (consisting in neglecting dominated
monomials and keeping only lowest order monomials in the differential equations).
This method leads to unbounded errors when fast cycles
are present in the reactions network. Indeed, tropical truncation can generate fast subsystems
that have conservation laws not present in the initial system.
Although this kind of truncation is accurate on short timescales,
it does not cope with slow relaxation of the mass carried by the fast cycles.
From a geometrical point of view, these conservation laws define sums of polynomials belonging
to the ideal and contribute to the
definition of the tropical variety. From a biochemical point of view, the conservation laws
provide pools of species whose total mass relaxes slowly and should stand as supplementary
slow variables. By this new approach, we use both pruning and pooling in order to reduce the
biochemical reactions networks.

\section{Newton-Puiseux series and tropical equilibrations.}
\label{newton}
In this section we introduce the Newton-Puiseux series and discuss their relation with
the tropical equilibrations.

By $K = \C((\epsilon^{1/\infty}))$ we denote the field of Newton-Puiseux series, i.e. all the series of the
type
\begin{equation}
x(\epsilon) = c_1 \epsilon^\frac{a_1}{q} + c_2 \epsilon^\frac{a_2}{q} +\ldots ,
\label{puiseux}
\end{equation}
where
$c_i \in \C$, $a_1 < a_2 < \ldots $ are integers, $q$ is a positive integer. The series
are convergent in some neighborhood of the origin, the origin being excluded if $a_1 <0$.

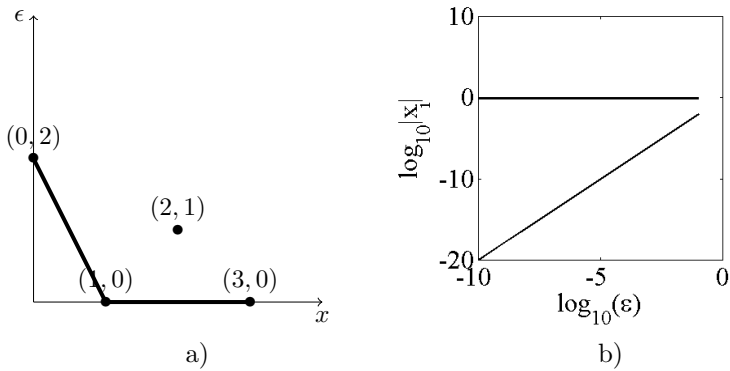
\begin{figure}[h]
\begin{center}
{\scalebox{0.95}{
\begin{tikzpicture}
  \draw [->,thin] (0, 0 ) -- (4, 0);
  \draw [->,thin] (0, 0 ) -- (0 , 4);
   \draw (4,0) node[below]{$x$} ;
   \draw (0,4) node[left]{$\epsilon$} ;
 \draw [ultra thick] (1, 0 ) -- (3, 0);
   \draw [ultra thick] (0, 2 ) -- (1, 0);
 \draw (3,0) node[above]{$(3,0)$} ;
 \draw (3,0) node{$\bullet$} ;
  \draw (1,0) node[above]{$(1,0)$} ;
   \draw (1,0) node{$\bullet$} ;
   \draw (2,1) node[above]{$(2,1)$} ;
    \draw (2,1) node{$\bullet$} ;
    \draw (0,2) node[above]{$(0,2)$} ;
     \draw (0,2) node{$\bullet$} ;
\end{tikzpicture}
}
}
\includegraphics[width=6cm]{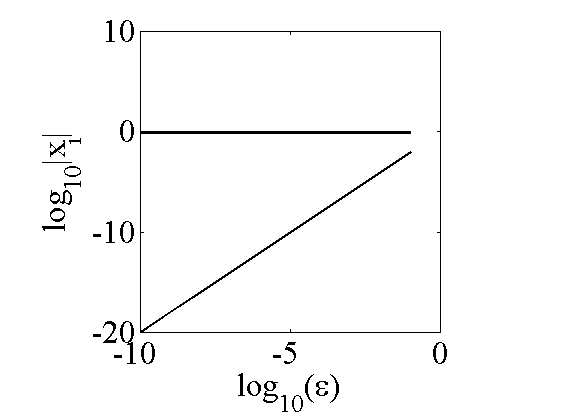}

a) \hskip5truecm b)
\end{center}
\caption{ \label{fig:newton} \small
Newton polygon and roots of the polynomial $x^3 + \epsilon x^2 - x + \epsilon^2$.
a) The Newton polygon edges are indicated by thick lines and the limiting monomials all
satisfy the sign condition.
The slopes of the edges are
$0$ and $-2$ corresponding to leading terms in the Newton-Puiseux series
of orders $\epsilon^0$ or $\epsilon^2$, respectively.
b) The absolute values of the polynomial's roots are represented vs. $\epsilon$, in logarithmic
scale; the slopes are the roots valuations.
}
\end{figure}

The Puiseux theorem \cite{Eisenbud} says that $K$ is
algebraically closed, i.e. that every nonconstant polynomial in $K[x]$ has a root in
$K$. In particular, any root of a polynomial whose coefficients are powers of $\epsilon$
can be written as a Newton-Puiseux series in $\epsilon$.
Furthermore, the leading term order $\frac{a_1}{q}$ can be calculated using the Newton polygon construction. Suppose we want to solve
the equation
\begin{equation}
P(x,\epsilon) = \sum_j k_j  \epsilon^{\gamma_j} x^{\alpha_j} =0,
\label{equ}
\end{equation}
where $\gamma_j$ are integers and $\alpha_j$ are positive integers.
In this case, Puiseux theorem ensures that Eq.\eqref{equ} has solutions
of the type \eqref{puiseux}. By substituting in \eqref{equ}
$x(\epsilon) = c_1 \epsilon^\frac{a_1}{q}(1 + x_1(\epsilon))$ (where $x_1(\epsilon)$ denotes terms
of order larger than zero in $\epsilon$)
we get
$$P(x,\epsilon) = \sum_j k_j  c_1^{\alpha_j} \epsilon^{\gamma_j + a_1 \alpha_j/q }  + r_1(\epsilon) =0,$$
where $r_1(\epsilon)$ collects higher order terms. Necessary conditions for $P(x,\epsilon)=0$
read at lowest order
\begin{eqnarray}
\sum_{j,\gamma_j + a_1 \alpha_j/q = m} k_j  c_1^{\alpha_j} = 0 \label{t1}\\
m = \min_j(\gamma_j + a_1 \alpha_j/q ) \label{t2}
\end{eqnarray}
In order to satisfy \eqref{t1}, the minimum in \eqref{t2}
should be attained at least twice. Furthermore, if one looks for
real solutions $c_1 \in \R$, then from \eqref{t1} it follows that
at least two $k_j$ corresponding to the minimum \eqref{t2} should
have opposite signs, namely:
\begin{equation}
min_{j,k_j >0}(\gamma_j + a_1 \alpha_j/q ) = min_{j,k_j <0}(\gamma_j + a_1 \alpha_j/q ).
\label{c1}
\end{equation}
 We should note that \eqref{c1} is a necessary,
but not sufficient condition for real solutions (for instance $x^2-x+1=0$
satisfies the condition but has no real solutions).
The above condition means that the lowest order $a_1/q$
in the Newton-Puiseux series solution has to satisfy
a tropical equilibration problem. Geometrically, $-a_1/q$ is the slope of
an edge of the Newton polygon, defined as the upper convex hull
of the points
of planar coordinates $(\alpha_j,\gamma_j)$ (i.e.
the convex hull including with any point the vertical half-line emanating up from this point).
For instance, the leading terms in solutions of  $x^3 + \epsilon x^2 - x + \epsilon^2 = 0$
have orders $\epsilon^0$ or $\epsilon^2$ (see Figure \ref{fig:newton}).
The Newton polygon method can be generalized to multivariate
polynomials and we have implemented it in an automatic algorithm for computing
tropical equilibrations presented elsewhere \cite{samal2014tropical}.

Fast variables of chemical reactions networks with multiple timescales satisfy quasi-stationarity equations. These are  multivariate polynomial equations of the type
\begin{equation}
P(x_1,x_2,\ldots,x_n, \epsilon) = \sum_j k_j  \epsilon^{\gamma_j} x_1^{(\alpha_j)_1}
x_2^{(\alpha_j)_2} \ldots x_n^{(\alpha_j)_n} =0,
\label{eq:kapranov}
\end{equation}
where $\gamma_j$ are integers and $(\alpha_j)_k$ are positive integers.
Let us note that $P(x) \in K[x_1,x_2,\ldots,x_n]$.
Like in the case of the univariate equation \eqref{equ}, the
tropical equilibrations provide lowest order
approximations of the solutions of \eqref{eq:kapranov}. We want to represent solutions
of \eqref{eq:kapranov} by series, in other words we want to lift the tropical solutions
to Newton-Puiseux series. In the univariate case, this is always possible by the
Puiseux theorem.
In the multivariate case, the lifting as
Newton-Puiseux series of any tropical solution is ensured by a theorem of Kapranov \cite{einsiedler2006non}.
In order to formulate this
result let us introduce the valuation function $Val(x)$ defined as the lowest power of
$\epsilon$ occurring in some Newton-Puiseux series $x(\epsilon)$. When applied to species
concentrations $x_i$, kinetic parameters $k_i$ and monomials $k_i x^{\alpha_i}$
the valuation gives the orders $a_i$, $\gamma_i$ and $\mu_i$, respectively.
As it can be easily checked
$$Val(x(\epsilon)) = \lim_{\epsilon \to 0} \log_{\epsilon} |x(\epsilon)|$$

Suppose that the  tropical equilibration condition defined in the introduction (Eq.\eqref{eq:minplus})
is satisfied.
%
By the same
method as in the univariate case it can be shown that \eqref{eq:minplus} is
a necessary condition to have real Newton-Puiseux solutions.
The Kapranov theorem \cite{einsiedler2006non} states that \eqref{eq:minplus} is also
a sufficient condition for having Newton-Puiseux solutions with prescribed
lowest orders $(a_1,a_2,\ldots,a_n)$. More precisely,
there are $x_i(\epsilon) \in K$ such that $Val(x_i)=a_i$ and such that $P(x_1(\epsilon),x_2(\epsilon),\ldots,x_n(\epsilon), \epsilon) =0$.

There is no analogue of Kapranov theorem working for systems of equations. In this case,
the condition \eqref{eq:minplus} though necessary, is not sufficient for guaranteeing
the existence of Newton-Puiseux solutions.

In general, in order to obtain tropical equilibrations that can be
lifted to Newton-Puiseux series we need to find a so-called tropical
basis \cite{bogart2007computing}. Let us consider that we want to find
approximate solutions of $n$ equations of the form
\eqref{eq:kapranov}, namely
\begin{equation}
P_1(\vect{x}, \epsilon)=0,\,P_2(\vect{x}, \epsilon)=0,\, \ldots, P_n(\vect{x}, \epsilon)=0
\label{systema}
\end{equation}
We first look for
vectors $(a_1,a_2,\ldots,a_n)\in \R^n$ satisfying the tropical
equilibration condition \eqref{eq:minplus} for each polynomial $P_k$, $k \in [1,n]$.
This set is included in the tropical prevariety.
Contrary to the case of one equation, it is not longer guaranteed that there are
solutions $x_1(\epsilon),x_2(\epsilon),\ldots, x_n(\epsilon)$
of \eqref{systema} such that
\begin{equation}
Val(x_1,x_2,\dots,x_n)=(a_1,a_2,\dots,a_n),
\label{val}
\end{equation}
where $Val$ means here application of the valuation componentwise.
Solutions of \eqref{systema} that satisfy \eqref{val}
can be found if we solve a more complex problem.
Let us supplement the system \eqref{systema}
with sums of products of the polynomials $P_1,P_2,\ldots,P_n$
by arbitrary polynomials
and
solve the tropical equilibration problem for the augmented system.
In other words, let us look for solutions in the tropical variety of the
ideal $I$ generated by $P_1,P_2,\ldots,P_n$.
By a result of \cite{speyer2004tropical} in this case there are Newton-Puiseux solutions that satisfy
the property \eqref{val}.
Although the ideal has an infinite number of elements, it can be shown
that it is enough to solve the tropical problem for a finite set of
polynomials in the ideal. A finite set of polynomials $f_1,f_2,\ldots,f_t$ generating
the ideal $I$ and such that $T(f_1) \cap T(f_2)\cap \ldots \cap T(f_t) = T(I)$
is called tropical basis.
An algorithm
for computing a tropical basis can be found in \cite{bogart2007computing}.
However, the complexity of this algorithm can be double-exponential in the size of the
system, both in time and in space.
In the remaining of this section we state a simple method
for finding tropical solutions that can
be lifted to Newton-Puiseux series.

Generically, a system of $n$ tropical equations in $n$ variables has a finite number of solutions.
Indeed, the equality of two
$n$-variate monomials corresponds to a $(n-1)$-dimensional hyperplane
in the space of coordinates $\log(x_i)$, or, equivalently, in the orders $a_i$.
The intersection of $n$ hyperplanes of dimension $(n-1)$ in a $n$-dimensional space is
generically a point. Because the combinations of
monomials that can be equilibrated are finite, the total number of solutions is finite.
However, chemical reactions systems often have infinite branches
of tropical equilibrations.
For instance, a reversible reaction
can equilibrate both reactants and reaction products. In quasi-equilibrium
conditions, the forward and reverse rate monomials of this reaction
are dominant, have equal orders and occur in equilibration equations
of several variables. Therefore, some of the hyperplanes
resulting from different tropical equations coincide. In these
cases, there are infinite sets of tropical equilibrations.

As an example, we can consider the following system of equations:
\begin{eqnarray}
y - x - \epsilon x^4 = 0 \notag \\
x - y +  \epsilon y^2 = 0
\label{system}
\end{eqnarray}
The valuations $a_1 = Val(x(\epsilon))$ and
$a_2 = Val(y(\epsilon))$ of
solutions of \eqref{system} satisfy
the tropical equations  $\min(a_1,1+4a_1)=\min(a_1,1+2a_2)=a_2$.
The condition $\min(a_1, 1 + 4a_1) =a_1$, leads to the infinite branch of
solutions $a_1=a_2 \geq -1/3$. The condition $\min(a_1, 1 + 4a_1) = 1 + 4a_1$ leads
to $\min(a_1, 1 + 2a_2)  = \min(a_1, 3 + 8a_1)  = 3 + 8a_1 = 1 + 4a_1$, therefore
 $a_1= -1/2$, $a_2= -1$, or to
$\min(a_1, 3 + 8a_1) =a_1 =1+4a_1$, hence $a_1=a_2=-1/3$ solution already found.
Thus the system \eqref{system} has an infinite branch of tropical equilbrations $a_1 = a_2 \geq -1/3$
and a isolated solution $a_1= -1/2$, $a_2= -1$.

However, only two of
these solutions lead to Newton-Puiseux solutions of \eqref{system}.
Indeed, the system \eqref{system} has $7$ complex solutions,
namely $(0,0)$, $(x, \pm x^2)$, where $x$ is a solution
of $\epsilon x^3 \mp x + 1 = 0$. Using the Newton polygon construction we
find that the possible valuations of
$x$ are $0$ or $-1/2$.
It follows that valuations of real
Newton-Puiseux solutions of Eq.~\eqref{system}
are $(0,0)$, or $(-1/2,-1)$. The only tropical solutions leading
to Newton-Puiseux series is $(0,0)$, a point on the continuous,
infinite branch of tropical solutions and $(-1/2,-1)$, the isolated
solution.


We conjecture that all the isolated tropical equilibrations
can be extended to Newton-Puiseux series.
Therefore, if by supplementing the original system with sums or products of the
original equations with arbitrary polynomials (i.e. considering the ideal generated by these equations)
we find a system with only isolated tropical equilibrations, we believe
that all of them can be lifted to Newton-Puiseux series.
For instance, in the above example, let us consider together to the equations \eqref{system} also their sum
$\epsilon(y^2 - x^4)=0$ and solve the resulting extended
tropical system $\min(a_1,1+4a_1)=\min(a_1,1+2a_2)=a_2, \quad a_1 = a_2/2$. This system has only two solutions,
$(0,0)$ and $(-1/2,-1)$. These two solutions are isolated. We have already shown that they can be lifted to Newton-Puiseux series.
An ansatz for finding linear combinations of equations
leading to isolated tropical equilibrations
is to consider conservations laws of the
fast subsystem. This ansatz will be used in
the Sections~\ref{general},\ref{simplecycle}.

\section{Model reduction of biochemical reaction newtworks with fast cycles.}
\label{general}
In this section we introduce our model reduction method. We also state and prove our main result on  the existence of invariant manifolds for networks with fast species and
fast cycles.

Let us call {\em tropically truncated system} associated to the tropical equilibration
$(a_1,a_2,\ldots,a_n)$, the system obtained by pruning all the dominated monomials
of \eqref{massaction} revealed
by the rescaling \eqref{massactionrescaled}, i.e.
\begin{equation}
 \D{x_i}{t} =   \sum_{j \in J(i)}   k_j S_{ij}   {{\vect{x}}}^{\vect{\alpha_{j}}}, \quad i \in [1,n]
 \label{massactiontruncated}
 \end{equation}
 where $J(i) = \underset{j}{\argmin} (\mu_{j},S_{ij}\neq 0)$ is the set of dominating
 reaction rates of reactions acting on species $i$ and $\mu_j$ are defined by \eqref{muj}.

Like in the introduction, we introduce the orders $\nu_i = \mu_{J(i)} - a_i$,
with $\mu_{J(i)} = \min(\mu_{j},S_{ij}\neq 0)$. The rescaled truncated
system reads
 \begin{equation}
 \D{\bar{x}_i}{t} = \varepsilon^{\nu_i} \sum_{j \in J(i)}  \bar k_j S_{ij}   {\bar{\vect{x}}}^{\vect{\alpha_{j}}},
 \quad i \in [1,n].
 \label{truncatedrescaled}
 \end{equation}
Variables $x_i$ with smaller orders $\nu_i$ are faster.
After reordering the variables we can consider that $\nu_1 \leq \nu_2 \leq \ldots \leq \nu_n$.
Let us assume that the following gap condition is fulfilled:
\begin{equation}
\text{there is } f<n \text{ such that } \nu_{f+1} - \nu_{f}  > 0,
\label{gapcondition}
\end{equation}
meaning that two groups of variables have separated timescales.
The variables $\vect{x}_f = (x_1,x_2,\ldots,x_f)$ are fast (change significantly on
timescales of order of magnitude $\epsilon^{-\nu_{f}}$ or shorter).
The remaining variables $\vect{x}_s = (x_{f+1},x_{f+2},\ldots,x_n)$ are
 slow (have little variation on timescales of order of magnitude $\epsilon^{-\nu_{f}}$).

 We call linear conservation law of a system of differential equations,
 a linear form $c(\vect{x}) = <\vect{c},\vect{x}> = c_1 x_1 + c_2 x_2 + \ldots + c_n x_n$
 that is identically constant on trajectories of the system.
It can be easily checked that vectors in the left kernel $Ker^{l}(S)$ of the stoichiometric matrix $\vect{S}$
provide linear conservation laws of the system \eqref{massaction}. Indeed,
system \eqref{massaction} reads $\D{\vect{x}}{t} = \vect{S} \vect{R}(\vect{x})$,
where the components of the vector $\vect{R}$ are $R_j(\vect{x}) = k_j x^{\alpha_j}$. If
 $\vect{c}^{T} \vect{S} = 0$, then $\D{<\vect{c},\vect{x}>}{t} = \vect{c}^{T} \vect{S} \vect{R}(\vect{x}) = 0$, where $\vect{c}^{T} =(c_1,c_2,\ldots,c_n)$.

Let us assume that the truncated system  \eqref{massactiontruncated}, restricted
to the fast variables has a number of independent,
linear conservation laws, defined by the vectors
$\vect{c}_1,\vect{c}_2,\ldots,\vect{c}_r$, where $\vect{c}_k = (c_{k1},c_{k2},\ldots,c_{kn})$.
These conservation laws can be calculated by recasting the truncated system
as the product of a new stoichiometric matrix and a vector of monomial rate
functions and further computing left kernel vectors of the new stoichiometric matrix.
We further assume that the truncated system and the full system \eqref{massactiontruncated} have no
conservation laws in common.

We now define the new variables
$y_k =  \sum_{l } c_{kl}  x_l$, where $k \in  [1,r]$.
These new variables satisfy the equations
\begin{equation}
\D{y_k}{t}  = \sum_{j }
\sum_{l } c_{kl} S_{lj} \vect{x}^{\vect{\alpha_j}}.
\label{ydyn}
\end{equation}
Let $(a_1,a_2,\ldots,a_n)$ be a solution of the tropical equilibration problem
for the augmented system obtained by putting together
\eqref{massaction} and \eqref{ydyn}. We define $b_k = \min(a_l, c_{kl} \neq 0 )$
and $\rho_k = \mu_{J_c(k)} - b_k$ where
$\mu_{J_c(k)} = \min (\mu_j, c_{kl} \neq 0, S_{lj} \neq 0)$,
$J_c(k) = \underset{j}{\argmin}  (\mu_j, c_{kl} \neq 0, S_{lj} \neq 0)$.
In rescaled variables $y_k = \bar{y}_k \epsilon^{b_k}$ we have the following truncated
rescaled system
\begin{equation}
\D{\bar y_k}{t}  = \epsilon^{\rho_k} \sum_{j \in J_c(k) }
\sum_{l } c_{kl} S_{lj} \bar{\vect{x}}^{\vect{\alpha_j}}.
\label{ydyntrunc}
\end{equation}
We assume that $\nu_f < \rho_k$, meaning that the variables $y_k, k \in [1,r]$ are slower than the variables
$x_i, i \in [1,r]$.

Since we have $r$ conservation laws, we can eliminate
$r$ fast variables from the truncated system.
One can suppose that these fast variables are
$x_{f-r+1}, ...,  x_f$ can be expressed via the
remaining variables $x_i$, $i\in[1,f-r]$ and $y_k$, $k\in [1,r]$.
Let us introduce vectors
$$
 \vect{X}_r=(\bar x_1,\bar x_2,\ldots, \bar x_{f-r}), \quad \vect{X}_s=(\bar x_{f+1},\bar x_{f+2}, \ldots, \bar x_n),
\quad \bar{\vect{y}} = (\bar y_{1}, \bar y_{2},\ldots, \bar y_r).
$$
Then any function of $\bar{\vect{x}}$ can be expressed via $\vect{X}_r, \vect{X}_s$ and $\bar{\vect{y}}$.
As a result, we obtain the following decomposition
\begin{equation}
 \D{\bar{x}_i}{t} = \varepsilon^{\nu_i} F_i(\vect{X}_r, \bar{\vect{y}}, \vect{X}_s, \epsilon)\
 \quad i \in [1,f-r],
 \label{truncatedrescaled5}
 \end{equation}
where
$$
F_i(\vect{X}_r, \bar{\vect{y}}, \vect{X}_s, \epsilon)=\sum_{j \in J(i)}  \bar k_j S_{ij}   {\bar{\vect{x}}}^{\vect{\alpha_{j}}},
$$
\begin{equation}
 \D{\bar{x}_i}{t} = \varepsilon^{\nu_i} S_i(\vect{X}_r, \bar{\vect{y}}, \vect{X}_s, \epsilon),
 \quad i \in [f+1,n],
 \label{truncatedrescaled6}
 \end{equation}
$$
S_i(\vect{X}_r, \bar{\vect{y}}, \vect{X}_s, \epsilon)=\sum_{j \in J(i)}  \bar k_j S_{ij}   {\bar{\vect{x}}}^{\vect{\alpha_{j}}},
$$
\begin{equation}
\D{\bar y_k}{t}  =\epsilon^{\rho_k}Y_k(\vect{X}_r, \bar{\vect{y}}, \vect{X}_s, \epsilon),
\quad k \in [1,r],
\label{ydyn2}
\end{equation}
$$
Y_k(\vect{X}_r, \bar{\vect{y}}, \vect{X}_s, \epsilon) = \sum_{j \in J_c(k) }
\sum_{l } c_{kl} S_{lj} \bar{\vect{x}}^{\vect{\alpha_j}},
$$
where $\vect{Y}$, $\vect{S}$ and $\vect{F}$ are analytic functions.

The system \eqref{truncatedrescaled5} describes the evolution of fast modes. Because it was
obtained from the truncated versions of the first $f-r$ equations of \eqref{massaction},
let us call it the fast truncated subsystem. As a matter of fact, the system \eqref{truncatedrescaled5} coincides with
the first $f-r$ equations of \eqref{truncatedrescaled}.

Let us recall some notions of the dynamical systems theory. Let $d\vect{x}/dt=\vect{F}(\vect{x})$ be a system of ordinary differential equations defined on an open domain $\Omega$ of an Euclidean space with smooth boundary. Here   $\vect{F}$ is a smooth function, for example,
 $\vect{F} \in C^r$, where $r > 1$. Let us consider an equilibrium (steady state) $\vect{\phi}$ (i.e., the relation $\vect{F}(\vect{\phi})=0$ holds)
of this system. Let $A$ be a linear operator  that gives a linearization of r.h.s. of this system  at  $\vect{\phi}$:
$$
         \vect{F}(\vect{x})= A (\vect{x} - \vect{\phi}) + O((\vect{x} - \vect{\phi})^2).
$$
We say that this equilibrium is hyperbolic \cite{Robinson}, if
the distance $d$ between  the spectrum $Spec_A$ of $A$ and the imaginary axis $I=\{z \in{\mathbb C} : Re \ z=0\}$
is not zero:
\begin{equation} \label{dist}
  d= dist( Spec A,  \  I ) \ne 0.
 \end{equation}
If the spectrum lies in the left-half plane $\{z \in{\mathbb C} : Re \ z < 0\}$, then
this equilibrium is stable and locally attracting.
In our  case all systems depend on the parameter $\epsilon > 0$, therefore,
$d$ in \eqref{dist} can depend on $\epsilon$.

We can now formulate our main result.

\begin{theorem}
Assume the gap condition \eqref{gapcondition} holds and that  $\nu_f < \rho_k$, $k \in [1,r]$.
Assume that for all values $\bar{\vect{y}}$ and $\vect{X_s}$ the fast truncated system \eqref{truncatedrescaled5} has
a stable hyperbolic steady state
$$\bar x_i = \phi_i ( \bar{\vect{y}}, \vect{X_s} ) + \text{higher order terms}, \quad i \in [1,f-r],$$ such that the distance $d(\epsilon)$ admits the estimate
$$
d > C_0 \epsilon^{\kappa}
$$
where $\kappa \ge 0$ is small enough and $C_0$ is independent on $\epsilon$.
Then  for sufficiently small $\epsilon >0$  system \eqref{massactiontruncated}
 has a locally attracting and locally invariant  normally hyperbolic $C^p$ ($p>1$) smooth manifold defined by
\begin{equation} \label{invman}
\bar x_i = \phi_i (\bar{\vect{y}}, \vect{X_s}, \epsilon) + \text{higher order terms}, \quad i \in [1,f-r],
\end{equation}
%
and the dynamics of the slow variables $\bar{\vect{y}}, \bar x_{f+1}, \bar x_{f+2}, \ldots, \bar x_n$ for large times takes the form
\begin{eqnarray}
 \D{\bar x_{i}}{t} =   \epsilon^{\nu_{i}} \sum_{j \in J(f+1)}   k_j S_{ij}   {{ \bar{\vect{x}}}_s}^{\vect{\alpha_{j}}}  + \text{higher order terms}
 ,\quad i \in [f+1,n] \notag \\
\D{\bar y_k}{t}  = \epsilon^{\rho_k} \sum_{j \in J_c(k) }
\sum_{l } c_{kl} S_{lj} \bar{\vect{x}}_s^{\vect{\alpha_j}} +  \text{higher order terms}, \quad k \in [1,r]
 \label{reduceddynamics}
 \end{eqnarray}
where $\bar{\vect{x}}_s^{\vect{\alpha_j}} = \phi_1^{\alpha_1^j}  \phi_2^{\alpha_2^j} \ldots \phi_f^{\alpha_f^j}  \bar x_{f+1}^{\alpha_{f+1}^j} \bar x_{f+2}^{\alpha_{f+2}^j}
\ldots  \bar x_{n}^{\alpha_n^j}$.
\label{bigtheorem}
\end{theorem}

{\bf Remark}:   ``higher order terms'' means  some smooth functions
of $\bar{\vect{y}}, \vect{X_s}, \epsilon$ having  higher orders in $\epsilon$
with respect to principal  terms.

{\em Proof.}
We  use the standard result, which follows from the theory of normally hyperbolic invariant manifolds \cite{Katok}. Consider  the system of differential equations
\begin{equation} \label{standard1}
  d\vect{u}/d\tau=A u + \lambda \vect{F}(\vect{u}, \vect{v}, \lambda) + \vect{H}(\vect{u}, \vect{v}, \lambda),
 \end{equation}
\begin{equation} \label{standard2}
  d\vect{v}/d\tau= \lambda^{\mu} \vect{S}(\vect{u}, \vect{v}, \lambda),  \quad \mu > 0,
 \end{equation}
 where $u \in {\mathbb R}^n$, $v \in {\mathbb R}^m$, $A$ is a linear operator
such that the spectrum of $A$ satisfies  (\ref{dist}) and lies in the left-half plane,
$F, G$ and $H$ are smooth functions uniformly bounded in $C^k$-norm on  ${\mathbb R}^n \times {\mathbb R}^m \times [0, 1]$ for some $k >1$. Moreover,
$H=O(|\vect{u}|^2)$ as $\vect{u} \to 0$.

 It is clear that
this system becomes slow/fast for small $\lambda >0$, where
$u$ are fast and $v$ are slow.  For any $p < k$ and sufficiently small
$\lambda$ there exists a normally hyperbolic smooth locally attracting invariant manifold close to $0$:
$u= \lambda U(v, \lambda)$, where $U$ is bounded in $C^p$ -norm.

To apply this result, we reduce our system  \eqref{massaction} to the form
\eqref{standard1}, \eqref{standard2}.
We introduce $ u= \vect{X_r} - \vect{\phi}$ and makes a time
change $\tau=\epsilon^{\kappa} t$.
We introduce the variables $v$ by $v=( \vect{X}_s, \bar{\vect{y}})$. Then,
if $\kappa>0$ is small enough we obtain that   system \eqref{truncatedrescaled5},
\eqref{truncatedrescaled6}, \eqref{ydyn2} can be rewritten
  in  the form \eqref{standard1}, \eqref{standard2} with
$\lambda=\epsilon^{\rho}$ for some $\rho >0$.
This completes the proof.
$\blacksquare$

\section{A simple nonlinear cycle example.}
\label{simplecycle}
Let us consider the following example of a cycle of reactions that includes a complex formation
reaction:
$$A_1 \overset{ k_1 }{\longrightarrow} A_2 \overset{ k_2 }{\longrightarrow} A_3
\overset{ k_3 }{\longrightarrow} A_1, \quad A_1  + A_2  \underset{k_5}{\overset{ k_4 }{\rightleftharpoons}}   A_3$$
The mass action chemical kinetic equations for this cycle read:
\begin{eqnarray}
\D{x_1}{t} & = & - k_1 x_1 + k_3 x_3 - k_4 x_1 x_2 + k_5 x_3 \label{cycle1_1} \\
\D{x_2}{t} &  =  & k_1 x_1 - k_2 x_2 - k_4 x_1 x_2 + k_5 x_3  \label{cycle1_2} \\
\D{x_3}{t} & =  & - k_3 x_3 + k_2 x_2 + k_4 x_1 x_2 - k_5 x_3
\label{cycle1_3}
\end{eqnarray}
Consider kinetic constants that scale like $k_i \sim \epsilon^{\gamma_i}$. For instance if $\epsilon = 1/10$
and $k_1=1, k_2=0.1, k_3=0.01, k_4=0.01, k_5=0.001$ we get
\begin{equation}
\gamma_1=0, \gamma_2=1, \gamma_3=\gamma_4=2, \gamma_5=3.
\label{orders}
\end{equation}

The tropical equilibrations for this model are the solutions of
  the following min-plus equations
\begin{equation}
\min ( \gamma_1 + a_1, \gamma_4 + a_1 + a_2 ) =  \min ( \gamma_3 , \gamma_5  ) + a_3 =
\min ( \gamma_2 + a_2, \gamma_4 + a_1 + a_2 ) =  \min ( \gamma_1 + a_1 , \gamma_5 + a_3 ),
\label{minpluscycle}
\end{equation}
where the first, second and third equality of \eqref{minpluscycle} follow from
\eqref{cycle1_1}, \eqref{cycle1_2}, \eqref{cycle1_3}, respectively.
Because $\gamma_3 < \gamma_5$ we have $\min ( \gamma_3 , \gamma_5  ) = \gamma_3$.
From \eqref{minpluscycle} it follows that  $\gamma_3  + a_3 = \min ( \gamma_1 + a_1 , \gamma_5 + a_3 )$. Furthermore $\min ( \gamma_1 + a_1 , \gamma_5 + a_3 ) =  \gamma_1 + a_1$ (because $\gamma_5 + a_3 >  \gamma_3 + a_3$).
Hence, in this case, the system of min-plus equations can be simplified to
\begin{equation}
\min ( \gamma_1 + a_1, \gamma_4 + a_1 + a_2 ) =   \gamma_3  + a_3 =
\min ( \gamma_2 + a_2, \gamma_4 + a_1 + a_2 ) =  \gamma_1 + a_1
\label{minpluscyclesimplified}
\end{equation}
Only one of the possible outputs of the first $\min$ operation
\eqref{minpluscyclesimplified}
has to be considered, namely
$\min ( \gamma_1 + a_1, \gamma_4 + a_1 + a_2 )=  \gamma_1 + a_1$, whereas the second $\min$
leads to two situations. It follows that there are thus two branches of tropical
solutions, namely:
\begin{eqnarray}
a_1 \geq \gamma_2 - \gamma_4, \quad a_2 &= a_1 + \gamma_1 - \gamma_2, \quad a_3 = a_1 + \gamma_1 - \gamma_3 \label{sol1} \\
a_1 \leq \gamma_2 - \gamma_4, \quad a_2 &= \gamma_1 - \gamma_4, \quad a_3 = a_1 + \gamma_1 - \gamma_3 \label{sol2}
\end{eqnarray}
The tropical equilibration solutions vary continuously on each of the branch and are non-isolated. It is thus possible that some tropical solutions or an entire branch can not be lifted to Newton-Puiseux
series. In order to find solutions that can be lifted we will consider equations in the ideal
of \eqref{cycle1_1}, \eqref{cycle1_2}, \eqref{cycle1_3}, starting with the conservation laws
of the fast subsystem. It will come out that this is enough, as adding these conservation laws
leads to isolated tropical equilibrations.

Let us first consider the  branch \eqref{sol1}.
By keeping dominating monomials of lowest order in $\epsilon$ and pruning all the others
we get the following truncated system:
\begin{equation}
(T)
\left\{
\begin{array}{lll}
\D{\bar x_1}{t} & =   & \epsilon^{\gamma_1} ( - \bar k_1 \bar x_1 + \bar k_3 \bar x_3 )    \\
\D{\bar x_2}{t} &  =  & \epsilon^{\gamma_2}  ( \bar k_1 \bar x_1 - \bar k_2 \bar x_2 ) \\
\D{\bar x_3}{t} & =   & \epsilon^{\gamma_3}  ( - \bar k_3 \bar x_3 + \bar k_2 \bar x_2 ),
\end{array}
\right.
\label{trop1}
\end{equation}
The truncated system \eqref{trop1} have the linear first integral
(conservation law)
\begin{equation}
y=x_1 + x_2 + x_3.
\label{cons}
\end{equation}
The variable $y$ is not a first integral of the full system,
which implies that the truncated system $(T)$ can not be a good approximation at large times.
The exact dynamics of $y$ is obtained by summing the equations \eqref{cycle1_1},\eqref{cycle1_2},\eqref{cycle1_3}:
\begin{equation}
\D{y}{t} = - k_4 x_1 x_2 + k_5 x_3.
\label{yeq}
\end{equation}
We consider that $y \sim \epsilon^{a_y}$ and further equilibrate the equations \eqref{cons},\eqref{yeq}. We therefore get
two more min-plus equations:
\begin{eqnarray}
a_y  =  \min (a_1,a_2,a_3) \label{newtrop1} \\
\gamma_4 + a_1 + a_2  =   \gamma_5 + a_3
\label{newtrop2}
\end{eqnarray}

Assume the particular choice \eqref{orders} of parameter orders.
Then, for the tropical solution \eqref{sol1}, it follows
$a_1 = 0$, $a_2=-1$, $a_y = a_3 = - 2$,
$\nu_1 = 0$, $\nu_2=1$, $\nu_3=2$, $\nu_y = 3$ which means that $y$ is slower than $x_i, 1 \leq i \leq 3$. The resulting tropical equilibration is in fact unique and thus isolated.
Indeed, considering the second branch of tropical equilibrations for the
variables $x_1,x_2,x_3$ \eqref{sol2} we find that $y$ can not be equilibrated because
\eqref{newtrop2} and \eqref{sol2} imply $\gamma_5 = \gamma_3$ which is not
satisfied. The polynomial defining the dynamics of
$y$ being in the ideal of polynomials defining the dynamics of
$x_1,x_2,x_3$, it follows that the branch \eqref{sol2} is not in the tropical variety and can
be discarded.


We will now use this isolated tropical equilibration to obtain reduced models.
We will discuss three reduced models: the truncated model $(T)$ in \eqref{trop1} that
describes the relaxation dynamics towards the attractive invariant manifold, the reduced model
 \eqref{reduceddynamics} given by Theorem \ref{bigtheorem} and describing
 dynamics on the invariant manifold, and a third reduced model combining these two.

The truncated
system \eqref{trop1} copes only with the fast relaxation onto the invariant manifold.
The tropical approximation of the invariant manifold is obtained by setting the l.h.s of
\eqref{trop1} to zero, i.e. computing the steady states of the truncated system. This
approximation is the half-line $x_2 = k_1 k_2^{-1} x_1, x_3 = k_1 k_3^{-1} x_1, x_1 < k_2 k_4^{-1}$.
By using the new tropically truncated equation:
\begin{equation}
y = x_3,
\label{tropcons}
\end{equation}
we compute $x_1$, $x_2$, $x_3$ from $y$ and obtain the reduced model:
\begin{eqnarray}
(R)
\left\{
\begin{array}{ll}
\D{y}{t} &= - k_1^{-1} k_2^{-1} k_3^2 k_4   y^2 + k_5 y  \\
x_1 &= k_1 ^{-1} k_3  y, \, x_2 = k_2^{-1} k_3 y, \, x_3 = y
\end{array}
\right.
\label{reduced}
\end{eqnarray}

The reduced model $(R)$ \eqref{reduced} copes with the slow dynamics on the invariant manifold.

In this particular example the two approximations $(T)$ and $(R)$ are composable, i.e.
they can be merged in a model with broader validity.
By replacing $y$ with $x_3$
in \eqref{reduced} and combining the resulting equations with the truncated system $(T)$ we
get the following model:
\begin{eqnarray}
(M)
\left\{
\begin{array}{lll}
\D{ x_1}{t} & =   & -k_1  x_1 +  k_3  x_3   \\
\D{ x_2}{t} &  =  &  k_1  x_1 -  k_2  x_2   \\
\D{x_3}{t} & = &  -k_3  x_3   +  k_2 x_2  - k_1^{-1} k_2^{-1} k_3^2 k_4   x_3^2 + k_5 x_3
\end{array}
\right.
\label{multiscale}
\end{eqnarray}

The model $(M)$ is a multiscale reduction as it gives accurate approximations of both fast and slow parts of the trajectories.

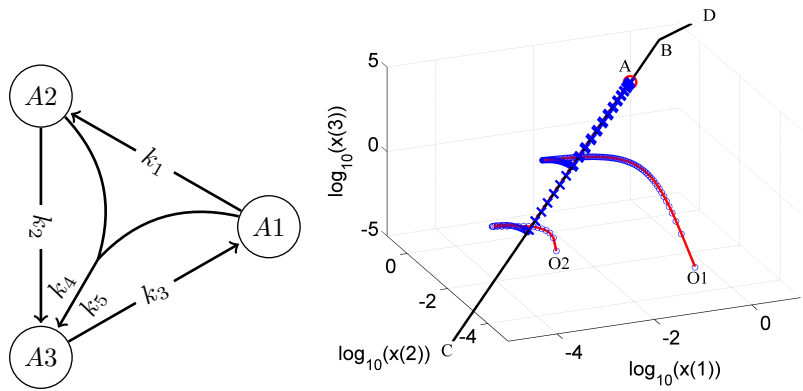
\begin{figure}[ht!]
\centering
\scalebox{1}{
\begin{tikzpicture}
 \SetUpEdge[lw         = 0.5pt,
            color      = black,
            labelstyle = {fill=white, sloped}]
  \tikzset{node distance = 2cm}
 \GraphInit[vstyle=Normal]
  \SetVertexMath
  \SetGraphUnit{2}
\Vertices{circle}{A1,A2,A3}
 \tikzset{EdgeStyle/.style={post,line width = 1}}
 \coordinate (A12) at ($(A1)!0.5!(A2)$) ;
 \coordinate (O) at ($(A12)!0.5!(A3)$) ;
 \Edge[label=$k_1$](A1)(A2)
 \Edge[label=$k_2$](A2)(A3)
 \Edge[label=$k_3$](A3)(A1)
 \tikzstyle{EdgeStyle}=[bend right,line width = 1]
  \Edge[](A1)(O)
  \tikzstyle{EdgeStyle}=[bend left,line width = 1]
 \Edge[](A2)(O)
  \SetUpEdge[lw         = 0.5pt,
            color      = black,
            labelstyle = {fill=none, sloped}]
 \tikzset{EdgeStyle/.style={post,line width = 1}}
 \Edge[label=$\genfrac{}{}{0pt}{0}{k_4}{k_5}$](O)(A3)
\end{tikzpicture}
}
\includegraphics[width=7cm]{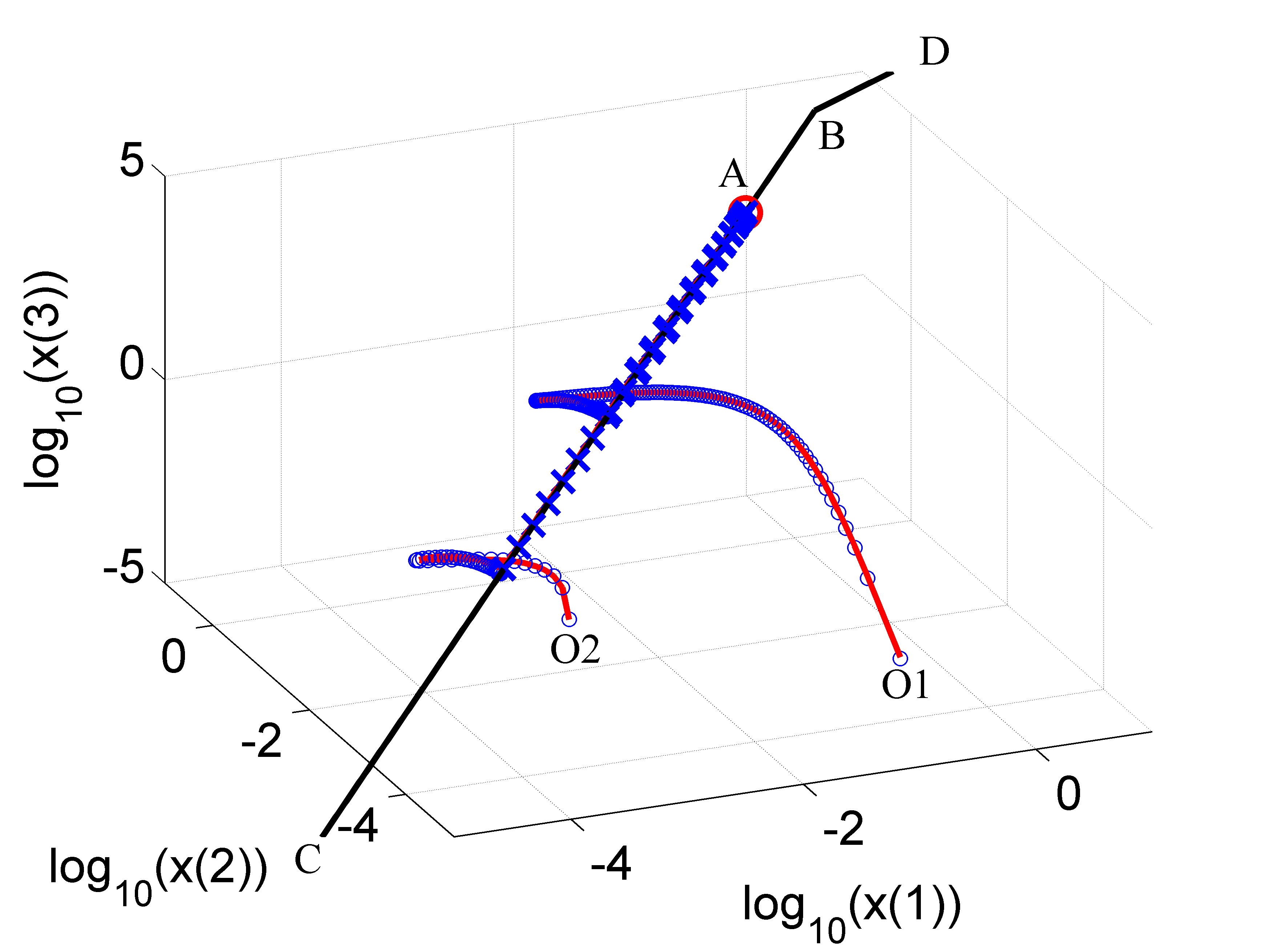}
 \caption{Two trajectories of the nonlinear cycle model
 defined by Eqs.\eqref{cycle1_1},\eqref{cycle1_2},\eqref{cycle1_3}
 starting from $O_1$ and $O_2$ are represented
 in red solid line. The blue circles are the trajectories starting from $O_1$ and $O_2$ computed
 with the tropical truncated model $(T)$ defined by Eqs.\eqref{trop1}. The blue crosses are the trajectories computed with the reduced
 model defined by Eqs.\eqref{reduced}.
 $A$ is the stable steady state of the model.
 The half-lines $BC$ and $BD$ belong to parts of tropical variety corresponding to the tropical solutions
 $a_1 \geq -1, \quad a_2 = a_1 -1 , \quad a_3 = a_1 -2 $
 and
 $a_1 \leq -1, \quad a_2 = -2, \quad a_3 = a_1 - 2$, respectivelly. }
\label{fig1}
\end{figure}

\begin{figure}[ht!]
\begin{center}
\includegraphics[width=8cm]{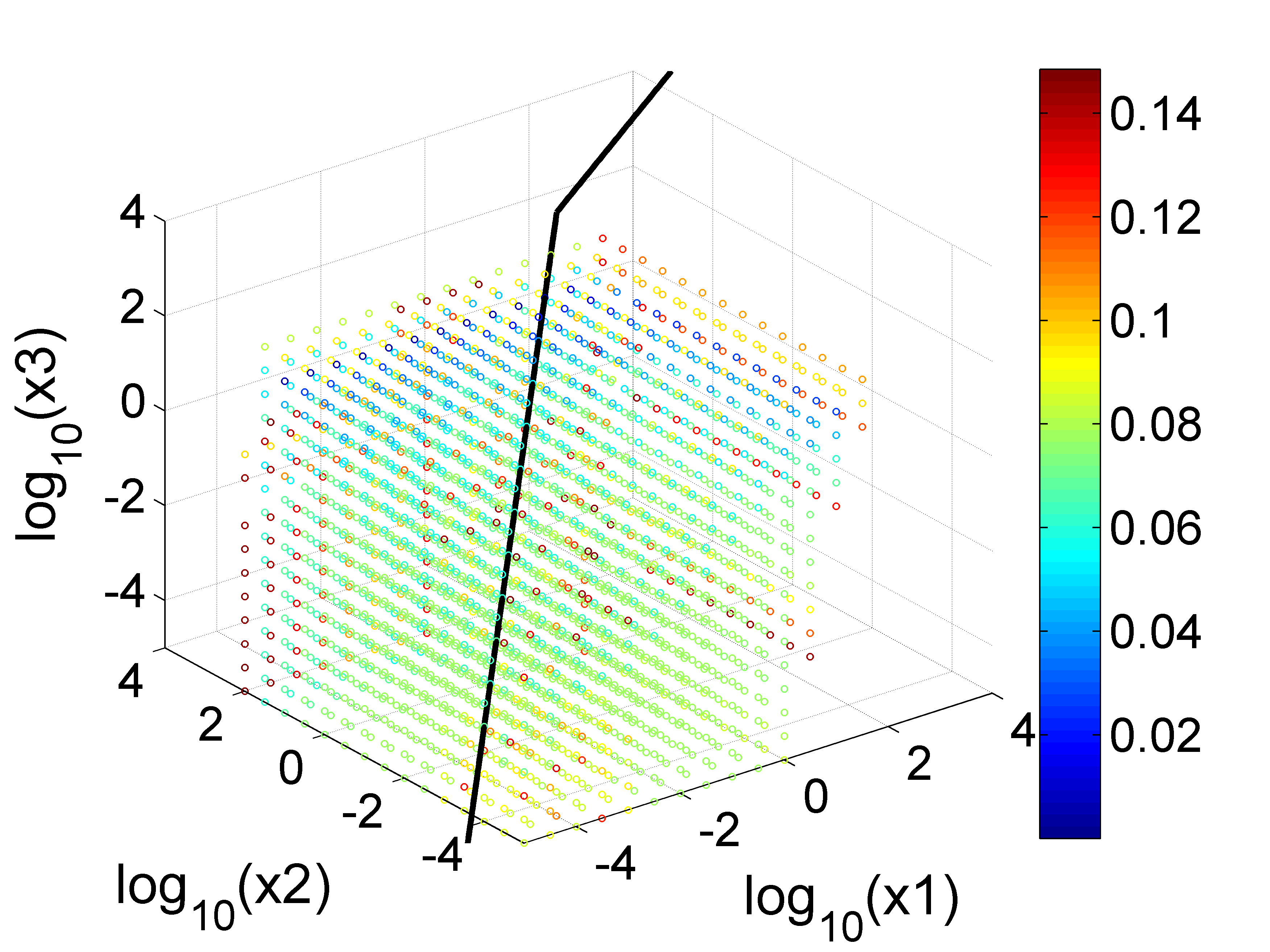}
\includegraphics[width=6cm]{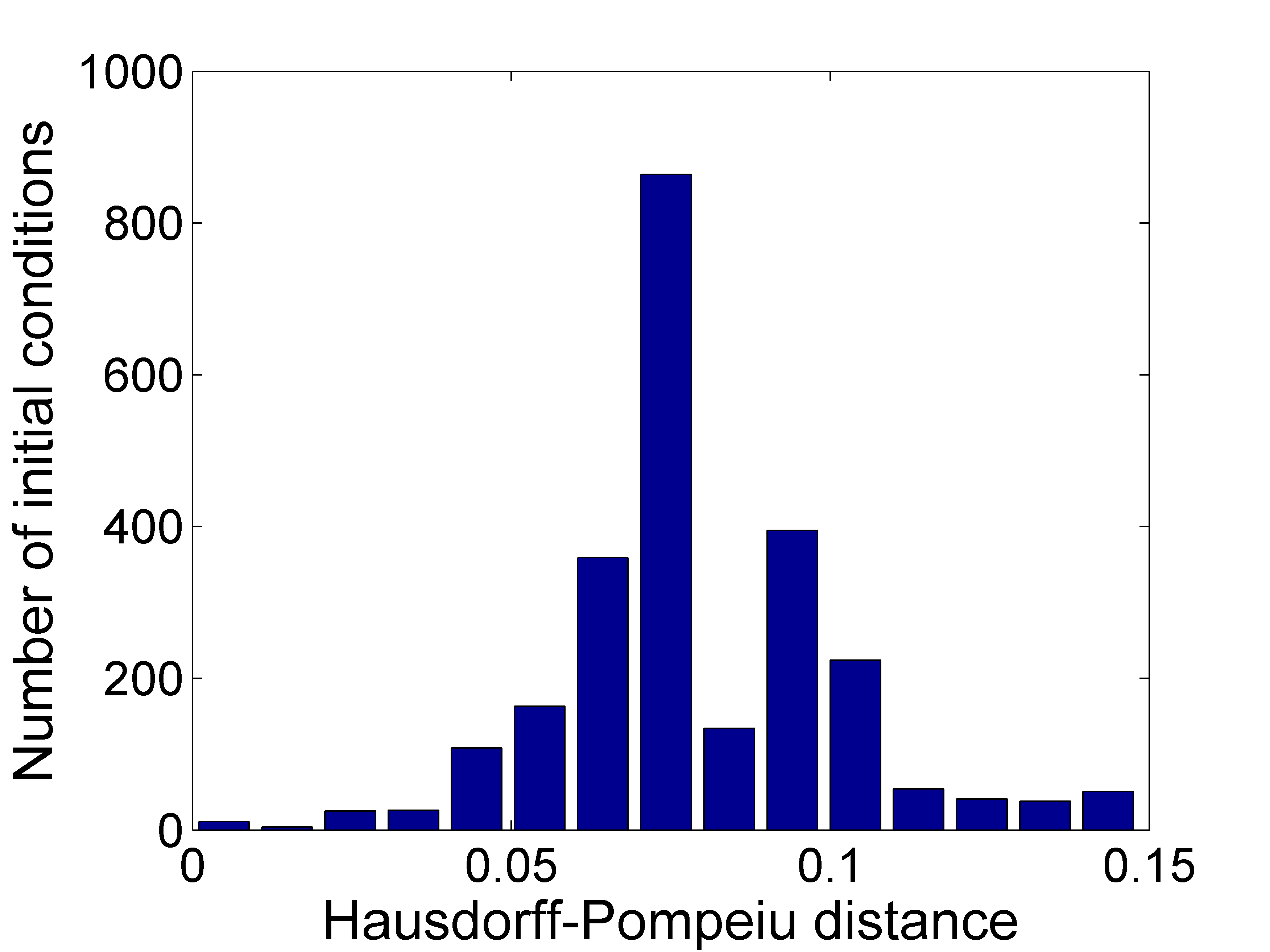}

a) \hskip5truecm b)
\end{center}
\caption{Trajectories of full (Eqs.\eqref{cycle1_1},\eqref{cycle1_2},\eqref{cycle1_3})  and multiscale reduced model (Eqs.\eqref{multiscale})  were computed starting from the same initial conditions and
the Hausdorff-Pompeiu distance between the corresponding sets
$\{ (log_{10}(t),log_{10}(x_1),log_{10}(x_2),log_{10}(x_3)), \text{variable}\, t\}$ were calculated. The initial conditions where chosen from an uniform grid in logarithmic scale
$-5 \leq log_{10}(x_1) \leq 4$,
$-5 \leq log_{10}(x_2) \leq 2$,
$-5 \leq log_{10}(x_3) \leq 4$.
a) The positions of the initial data leading to Hausdorff-Pompeiu distance less than $0.15$ are shown by circles with color coded values of this distance. The lines indicates
the same parts of tropical variety as in Fig.\ref{fig1}. Initial data can vary on $7$ decades
with global relative errors less than $1-10^{0.15} \approx 0.4$  which for $\epsilon =1/10$
and stiff trajectories is remarkably robust.
b) The distribution of Hausdorff-Pompeiu distances is shown for the set of initial data.
}
\label{fig3}
\end{figure}

The comparison
of different approximations is shown in Figure~\ref{fig1}.
The validity of the multiscale reduction could depend on the initial data.
We have determined numerically the domain of initial data leading
to accurate reductions. Starting from the same initial data we have
integrated the full model \eqref{cycle1_1},\eqref{cycle1_2},\eqref{cycle1_3} and
the reduced model \eqref{multiscale} and obtained
the trajectories $x(t)$ and $xr(t)$, respectively. We have computed the error
such as Hausdorff-Pompeiu distance between the sets $\{ (log_{10}(t),log_{10}(x_1(t)),log_{10}(x_2(t)),log_{10}(x_3(t))), \text{variable}\, t\}$ and $\{ (log_{10}(t),log_{10}(xr_1(t)),log_{10}(xr_2(t)),log_{10}(xr_3(t))), \text{variable}\, t \}$.  We notice in Figure~\ref{fig3}a)
that we can change the initial data on $7$ decades and still keep the trajectories
of the reduced model \eqref{multiscale} close to the trajectories of
the full model \eqref{cycle1_1},\eqref{cycle1_2},\eqref{cycle1_3}.

%
%
%

\section{Conclusion}
We have shown how to relate the tropical equilibration problem to the slow-fast
decomposition and model reduction of biochemical reactions networks.
In the case of biochemical networks with mass action kinetics, we use
tropical equilibration solutions to find which species are fast and which
are slow.
We have proposed elsewhere two methods for solving the tropical equilibration
problem, a first one by reformulating it as a constraint satisfaction problem
\cite{soliman2014constraint} and a second one based on the Newton
polyhedron \cite{samal2014tropical}.

Under rather general conditions, existence
of small dimensional attractive invariant manifolds for reactions networks with
fast cycles and species is shown.

Our model reduction recipe consists in calculating tropical equilibration
solutions at least twice. At the first step we solve the tropical equilibration
problem for the initial system of differential equations.
This allows us to identify the fast species, that constitute the fast subsystem
of the model. The fast truncated system, obtained by pruning dominated
monomials in the equations of fast species, copes with relaxation towards
the attracting invariant manifold.
The tropical equilibrations calculated at this step belong to the tropical prevariety,
but not all of them lead to a reduced model. In order to filter tropical solutions
we solve the tropical equilibration problem a second time.
If the fast truncated subsystem has conservation laws different from the
ones of the full system,
we use them to define new slow variables.
 At the second step, we solve
the tropical equilibration problem
 for the augmented system that is
obtained by adding to the initial system the differential equations satisfied by the conservation laws of the
fast truncated subsystem. These new equations are linear combinations
of the initial ones. We conjecture that if the resulting solutions are isolated, then
they belong to the tropical variety. Also, they lead to reduced models obtained by
expressing fast variables in terms of the slow variables. The resulting
reduced model copes with the dynamics on the invariant manifold.
If after the second step one still has truncated systems with conservation laws and
continuous branches of tropical equilibrations, the first two steps can be reiterated
until there are no conservation laws different from the ones of the full model and all tropical equilibrations are isolated.

Our method can be used as recipe for formal model reduction
in computational biology.
Some steps of the recipe are already automated, such as the calculation of tropical equilibrations. The computation of conservation laws can be performed by methods from \cite{soliman2012invariants}, but we don't exclude the existence of difficult cases
when conservation laws are not enough for grasping the tropical variety.
In these difficult cases, calculation of tropical basis can use methods from
\cite{bogart2007computing}.
Another challenging step is the elimination of fast variables as solutions of systems of polynomial equations. When the polynomials of the fast truncated system contain
only two monomials, we can apply rapid methods for toric
systems \cite{millan2012chemical,grigoriev2012}.
In general, fast truncated systems are fewnomials (have a small number of monomials)
and can be approached by the methods for sparse polynomial systems \cite{grigoriev2012}. However, even with fewnomials, there
could be models for which the elimination of fast variables is difficult.
Numerical methods can be used, as a last resort, to solve
problematic cases.

For a simple example, we suggested, without providing a general recipe,
how to combine one-scale approximations to obtain a multiscale approximation
that is valid on both fast and slow time scales. A general method for
obtaining multiscale approximations was given in \cite{gorban-dynamic}
for networks of monomolecular reactions (in these networks
each reaction has at most one reactant and at most one product
and the reaction rates are given by the mass action law) with
separated kinetic constants. Multiscale approximations of nonlinear networks
are much more difficult and will be discussed elsewhere.

\section*{Acknowledgements}
O. Radulescu has been supported by the PEPS CNRS ModRedBio
and Labex EPIGENMED (ANR-10-LABX-12-01). D. Grigoriev is grateful to the
Max-Planck Institut f\"ur Mathematik, Bonn for its hospitality
during writing this paper and to  Labex CEMPI (ANR-11-LABX-0007-01).
S. Vakulenko has been supported in part by Linkoping University,
by Government of Russian Federation, Grant 074-U01 and by grant 13-01-00405-a
of Russian Fund of Basic Research. Also he was supported in part by grant
RO1 OD010936 (formerly RR07801) from the US NIH.


\end{document}